\documentclass[prl,superscriptaddress,twocolumn,aps,10pt]{revtex4-1}
\pdfoutput=1
\usepackage[colorlinks=true,urlcolor=blue]{hyperref}
\usepackage{amsfonts}
\usepackage{graphicx}
\usepackage{mathrsfs}
\usepackage{amsmath}
\usepackage{xcolor}
\usepackage{bm}
\usepackage{upgreek}

\usepackage{color}
\definecolor{red}{rgb}{0.75,0,0}
\definecolor{blue}{rgb}{0,0,0.75}
\definecolor{green}{rgb}{0,0.5,0}

\begin{document}
\title{Cytoskeletal Anisotropy Controls Geometry and Forces of Adherent Cells} 

\author{Wim Pomp}
\thanks{W.P. and K.S. contributed equally and are the primary authors of this work.}
\affiliation{Kamerlingh Onnes-Huygens Laboratory, Leiden University, Niels Bohrweg 2, 2333 CA, Leiden,  Netherlands}
\author{Koen Schakenraad}
\thanks{W.P. and K.S. contributed equally and are the primary authors of this work.}
\affiliation{Instituut-Lorentz, Leiden University, P.O. Box 9506, 2300 RA Leiden, Netherlands}
\affiliation{Mathematical Institute, Leiden University, P.O. Box 9512, 2300 RA Leiden, Netherlands}
\author{Hayri~E.~Balc{\i}o\u{g}lu}
\affiliation{Toxicology, Leiden Academic Center for Drug Research, Leiden University, Netherlands}
\author{Hedde~van~Hoorn}
\affiliation{Kamerlingh Onnes-Huygens Laboratory, Leiden University, Niels Bohrweg 2, 2333 CA, Leiden, Netherlands}
\author{Erik~H.~J.~Danen}
\affiliation{Toxicology, Leiden Academic Center for Drug Research, Leiden University, Netherlands}
\author{Roeland M. H. Merks}
\affiliation{Mathematical Institute, Leiden University, P.O. Box 9512, 2300 RA Leiden, Netherlands}
\affiliation{Institute of Biology, Leiden University, P.O. Box 9505, 2300 RA Leiden, Netherlands}
\author{Thomas Schmidt}
\affiliation{Kamerlingh Onnes-Huygens Laboratory, Leiden University, Niels Bohrweg 2, 2333 CA, Leiden, Netherlands}
\author{Luca Giomi}
\thanks{Corresponding author: giomi@lorentz.leidenuniv.nl}
\affiliation{Instituut-Lorentz, Leiden University, P.O. Box 9506, 2300 RA Leiden, Netherlands}

\begin{abstract}
We investigate the geometrical and mechanical properties of adherent cells characterized by a highly anisotropic actin cytoskeleton. Using a combination of theoretical work and experiments on micropillar arrays, we demonstrate that the shape of the cell edge is accurately described by elliptical arcs, whose eccentricity expresses the degree of anisotropy of the internal cell stresses. This results in a spatially varying tension along the cell edge, that significantly affects the traction forces exerted by the cell on the substrate. Our work highlights the strong interplay between cell mechanics and geometry and paves the way towards the reconstruction of cellular forces from geometrical data.
\end{abstract}

\maketitle

Cells, from simple prokaryotes to the more complex eukaryotes, are capable of astonishing mechanical functionalities. They can repair wounded tissues by locally contracting the extra-cellular matrix \cite{Midwood2004}, move in a fluid or on a substrate \cite{Barry2010}, and generate enough force to split themselves in two while remaining alive \cite{Tanimoto2012}. Conversely, cell behavior and fate crucially depend on mechanical cues from outside the cell~\cite{Geiger2009,Discher2005,Julicher2007,Hoffman2011,Mendez2012}. Examples include rigidity-dependent stem cell differentiation \cite{Engler2006,Trappmann2012}, protein expression regulated by internal stresses \cite{Sawada2006}, mechanical cell-cell communication \cite{King2008} and durotaxis \cite{Lo2000,Sochol2011}. In all these biomechanical processes, cells rely on their shape \cite{Bischofs2009,Ghibaudo2009,Fletcher2010} to gauge the mechanical properties of their microenvironment \cite{Balcioglu2015} and direct the traction forces exerted on their surroundings. 

\begin{figure*}[t]
\centering
\includegraphics[width=\textwidth]{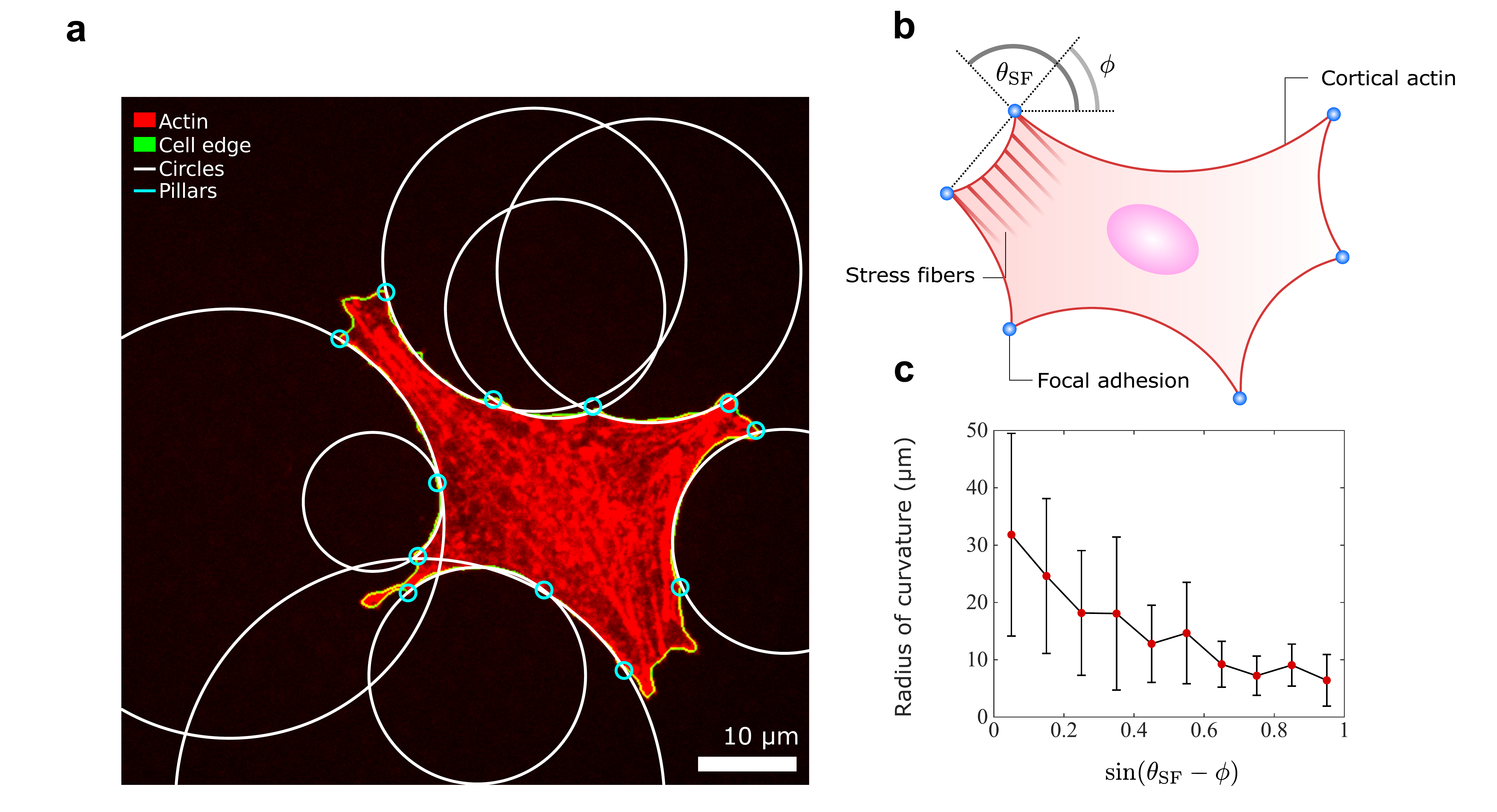}
\caption{\label{fig1}(a) A cell with an anisotropic actin cytoskeleton (epithelioid GE$\upbeta$3) with circles (white) fitted to its edges (green). The end points of the arcs (cyan) are identified based on the forces exerted on the pillars \cite{Sup}. The actin cytoskeleton is visualized with tetramethyl isothiocyanate rhodamine phalloidin (red). Scalebar is 10 $\upmu$m. (b) The cell cortex (red line) is spanned in segments between fixed adhesion sites (blue). (c) Arc radius as a function of the sine of the angle $\theta_{\rm SF}-\phi$, between the local orientation of the stress fibers and that of the distance between the adhesion points (data correspond to a sample of 285 cells and show the mean $\pm$ standard deviation).}
\end{figure*} 
 
In recent years, experiments on adhesive surfaces have contributed to explore such mechanical complexity in a controlled setting \cite{Schwarz2013}. Immediately after coming into contact with such a surface, many animal cells spread and develop transmembrane adhesion receptors. This induces the actin cytoskeleton to reorganize into cross-linked networks and bundles (i.e., stress fibers \cite{Pellegrin2007,Burridge2013}), whereas adhesion becomes limited to a number of sites, distributed mainly along the cell contour (i.e., focal adhesions \cite{Burridge1996}). At this stage, cells are essentially flat and assume a typical shape characterized by arcs which span between the sites of adhesion, while forces are mainly contractile \cite{Schwarz2013}. On timescales much shorter than those required by a cell to change its shape (i.e., minutes), the cell can be considered in mechanical equilibrium at any point of its interface. These observations have opened the way to the use of theoretical concepts inspired by the physics of fluid interfaces \cite{Bar-Ziv1999,Bischofs2008,Bischofs2009,Schwarz2013}, but limited to the case of cells with an isotropic cytoskeleton.

In this Letter, we overcome this limitation and explore the geometry and the mechanical properties of adherent cells characterized by a highly anisotropic actin cytoskeleton. Using a combination of theoretical modeling, spinning disk confocal microscopy, and traction-force microscopy of living cells cultured on microfabricated elastomeric pillar arrays \cite{Tan2003,Trichet2012a,VanHoorn2014}, we demonstrate that both the shape of and the traction forces exerted by adherent cells are determined by the anisotropy of their actin cytoskeleton. In particular, by comparing different cell types \cite{Danen2002}, we demonstrate that the cell contour consists of arcs of a unique ellipse, whose eccentricity expresses the degree of anisotropy of the internal stresses.

\begin{figure*}[t]
\centering
\includegraphics[width=\textwidth]{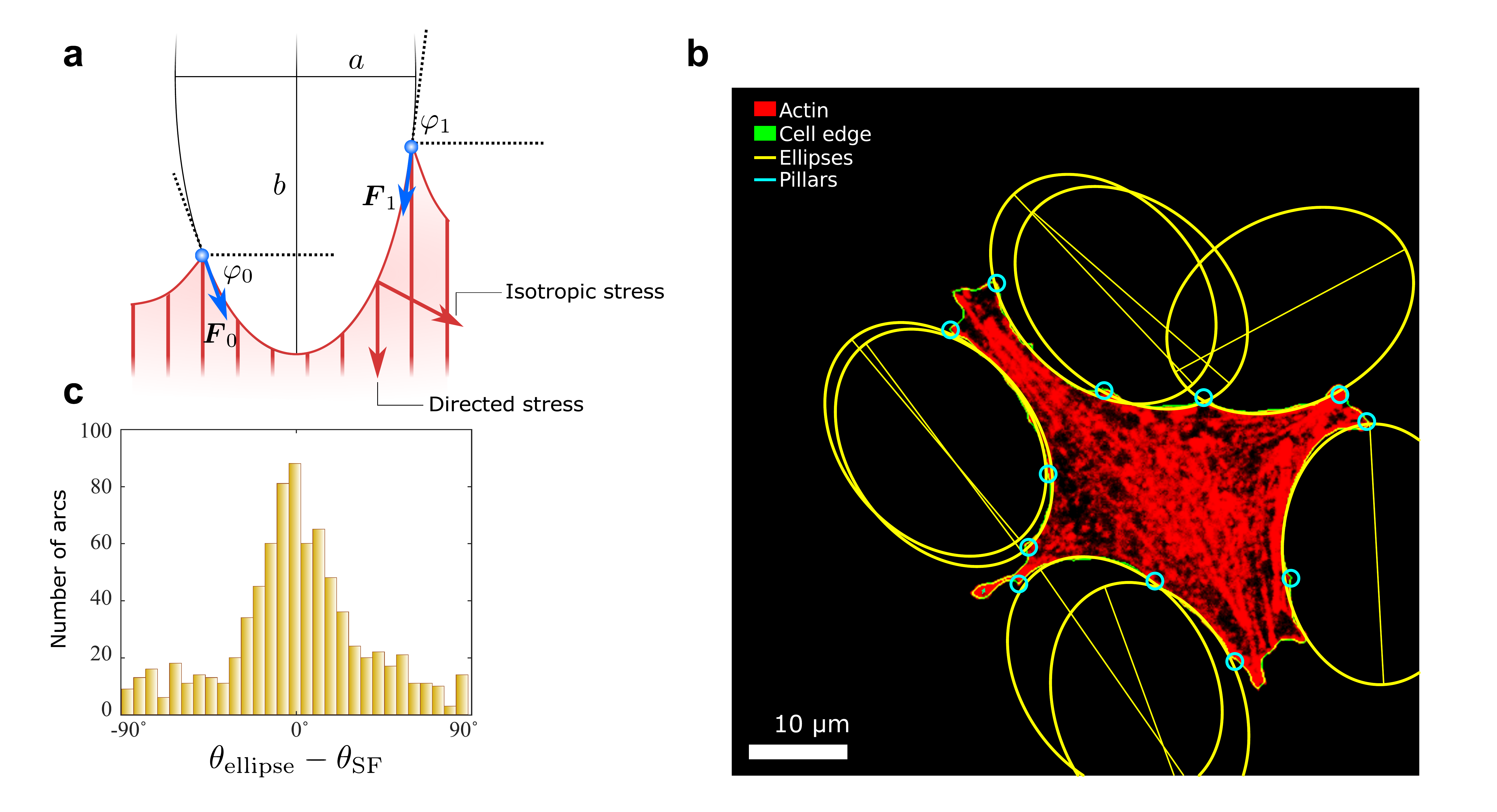}
\caption{\label{fig2}(a) Schematic representation of our model for $\theta_{\rm SF}=\pi/2$. All cellular arcs are part of a unique ellipse of aspect ratio $a/b=\sqrt\gamma$. The cell exerts forces $\bm F_0$ and $\bm F_1$ on the adhesion sites (blue) with magnitude $\lambda(\varphi_0)$ and $\lambda(\varphi_1)$. (b) An epithelioid cell (GE$\upbeta$3; same cell as in Fig.~\ref{fig1}a) with a unique ellipse (yellow) fitted to its edges (green). The end points of the arcs (cyan) are identified based on the forces exerted on the pillars \cite{Sup}. The fitted values of the ellipses' major and minor axes are, respectively, $13.38\pm 0.04\,\upmu{\rm m}$ and $9.65\pm 0.02\,\upmu{\rm m}$. The major axes (yellow lines) are parallel to the stress fibers. Their orientations are found to be, in counterclockwise order from the nearly vertical ellipse in the bottom right corner, $\theta_{\rm SF}=93\pm 4^{\circ},\,28\pm 5^{\circ},\,110\pm 2^{\circ},\,139\pm 6^{\circ},\,127\pm 3^{\circ},\,125\pm 2^{\circ},\,133\pm 2^{\circ},\,130\pm 3^{\circ}$ with respect to the horizontal axis of the image. Scalebar is 10 $\upmu$m. (c) Histogram of $\theta_{\rm ellipse}-\theta_{\rm SF}$, with $\theta_{\rm ellipse}$ the orientation of the major axis of the fitted ellipse and $\theta_{\rm SF}$ the measured orientation of the stress fibers. The mean of this distribution is $0^{\circ}$ and the standard deviation is $36^{\circ}$.}
\end{figure*} 
 
We model adherent cells as two-dimensional contractile films \cite{Giomi:2013a,Giomi:2013b}, and we focus on the shape of the cell edge connecting two consecutive adhesion sites. Mechanical equilibrium requires the difference between the internal and external stresses acting on the cell edge to balance the contractile forces arising in the cortex:
\begin{equation}\label{LP}
\frac{{\rm d}\bm{F}_{\rm cortex}}{{\rm d}s}+(\bm{\hat\Sigma}_{\rm out}-\bm{\hat\Sigma}_{\rm in}) \cdot \bm{N} = \bm{0}\;.
\end{equation} 
Here $\bm{\hat\Sigma}_{\rm out}$ and $\bm{\hat\Sigma}_{\rm in}$ are the stress tensors outside and inside the cell and $\bm{F}_{\rm cortex}$ is the stress resultant along the cell cortex. The latter is parametrized as a one-dimensional curve spanned by the arc-length $s$ and oriented along the inward pointing normal vector $\bm{N}$. A successful approach, initially proposed by Bar-Ziv {\em et al}. in the context of cell pearling \cite{Bar-Ziv1999} and later expanded by Bischofs {\em et al}. \cite{Bischofs2008,Bischofs2009}, consists of modeling bulk contractility in terms of an isotropic pressure $\bm{\hat\Sigma}_{\rm out}-\bm{\hat\Sigma}_{\rm in}=\sigma\bm{\hat I}$, with $\bm{\hat I}$ the identity matrix, and peripheral contractility as an interfacial tension of the form $\bm{F}_{\rm cortex}=\lambda \bm{T}$, with $\bm{T}$ a unit vector tangent to the cell edge. The quantities $\sigma$ and $\lambda$ are material constants that embody the biomechanical activity of myosin motors in the actin cytoskeleton. This competition between bulk and peripheral contractility along the cell boundary results in the formation of arcs of constant curvature $1/R = \sigma/\lambda$, through a mechanism analogous to the Young-Laplace law for fluid interfaces. The shape of the cell boundary is then approximated by a sequence of circular arcs, whose radius $R$ might or might not be uniform across the cell, depending on how the cortical tension $\lambda$ varies from arc to arc. The possibility of an elastic origin of the cortical tension was also explored in Ref. \cite{Bischofs2008} to account for an apparent correlation between the curvature and length $L$ of the cellular arcs. In this case $\lambda=k (L-L_{0})/L_{0}$, with $k$ an elastic constant and $L_{0}$ a rest length. Both models successfully describe the geometry of adherent cells in the presence of strictly isotropic forces.

Yet, many cells, including the fibroblastoids (GD$\upbeta$1, GD$\upbeta$3) and epithelioids (GE$\upbeta$1, GE$\upbeta$3) \cite{Danen2002} studied here [Fig.~\ref{fig1}(a)], develop directed forces by virtue of the strong anisotropic cytoskeleton originating from the actin stress fibers \cite{Pellegrin2007,Burridge2013}. This scenario is, evidently, beyond the scope of models based on isotropic contractility. Indeed, long cellular arcs appear prominently non-circular, as indicated by the fact that their curvature smoothly varies along the arc up to a factor ten [Fig. S1(a) in the Supplemental Material  \cite{Sup}]. Furthermore, whereas the shape of the cell edge in Fig.~\ref{fig1}(a) can in principle be approximated by circular arcs, a survey of a sample of 285 cells [Fig. S1(b) in the Supplemental Material \cite{Sup}] did not allow conclusive statements about a possible correlation between the arcs length and curvature, required to justify the variance in $\lambda$ \cite{Bischofs2008,Bischofs2009}. On the other hand, our data show a significant correlation between the radius of curvature of the cellular arcs and their orientation with respect to the stress fibers [Fig.~\ref{fig1}(b)]. In particular, the radius of curvature decreases as the stress fibers become more perpendicular to the cell cortex [Fig.~\ref{fig1}(c)]. This correlation is intuitive as the bulk contractile stress focusses in the direction of the stress fibers.

The anisotropy of the actin cytoskeleton can be incorporated into the mechanical framework summarized by Eq. \eqref{LP}, by modeling the stress fibers as contractile force-dipoles. This collectively gives rise to a directed contractile bulk stress, such that $\bm{\hat\Sigma}_{\rm out}-\bm{\hat\Sigma}_{\rm in}=\sigma\bm{\hat I}+\alpha\bm{n}\bm{n}$ \cite{Pedley1992,Simha2002,Giomi:2013c}, with $\bm{n}=(\cos\theta_{\rm SF},\sin\theta_{\rm SF})$ the average direction of the stress fibers [Fig. 1(b)]. The quantity $\alpha>0$ represents the magnitude of the directed contractile stresses and is proportional to the local degree of alignment between the fibers. The higher the alignment, the larger $\alpha$, whereas in the case of randomly oriented fibers $\alpha=0$, thus recovering the isotropic case. The ratio between isotropic contractility $\sigma$ and directed contractility $\alpha$ measures the degree of anisotropy of the bulk stress. With this stress tensor the force balance Eq. \eqref{LP} becomes
\begin{equation}\label{shape}
\frac{{\rm d}\lambda}{{\rm d}s}\,\bm{T}+(\lambda\kappa+\sigma)\bm{N}+\alpha(\bm{n}\cdot\bm{N})\bm{n}=\bm{0}\;,
\end{equation}
where we use $\mathrm{d}\bm{T}/\mathrm{d} s = \kappa \bm{N}$, with $\kappa$ the curvature of the cell edge. This implies that, in the presence of an anisotropic cytoskeleton, the cortical tension $\lambda$ is no longer constant along the cell cortex, as long as the directed stress has a non-vanishing tangential component (i.e. $\bm{n}\cdot\bm{T} \ne 0$). As shown by Kassianidou {\em et al}. \cite{Kassianidou:2017}, isolated stress fibers can also exert localized contractile forces on the cell contour, leading to kinks and piecewise constant curvature. Consistent with our experiments, here we consider the case in which the density of the stress fibers is sufficiently high and uniform to approximate their mechanical effect in terms of a continuous anisotropic stress.

\begin{figure*}[t]
\centering
\includegraphics[width=\textwidth]{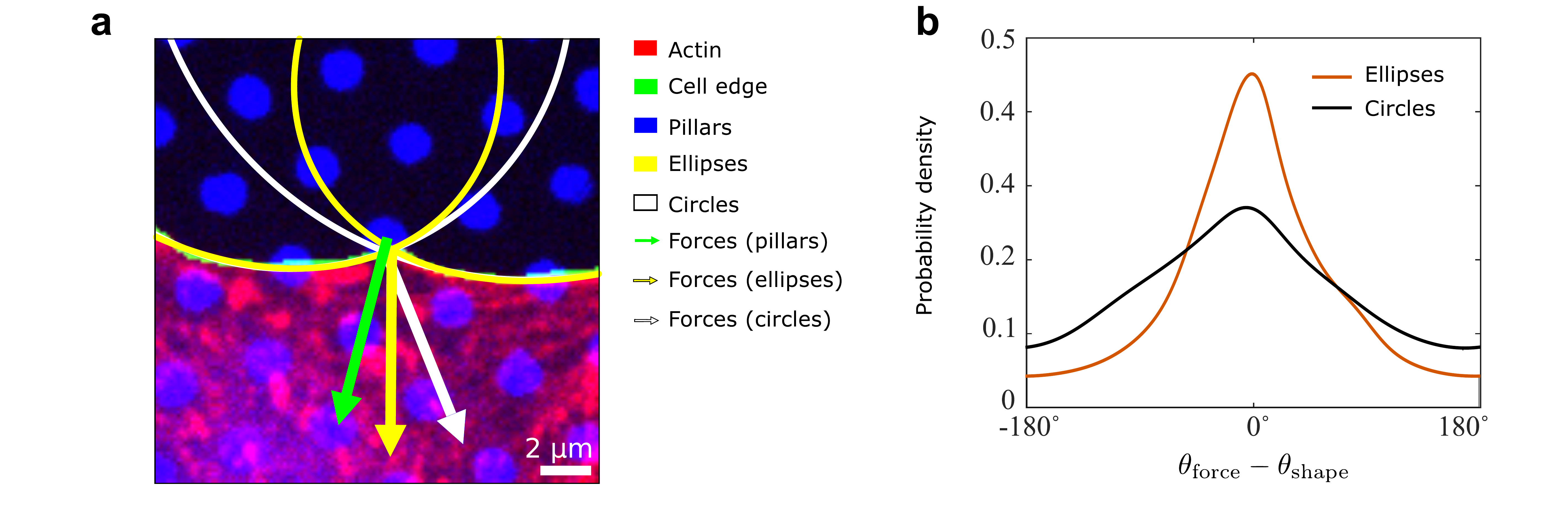}
\caption{\label{fig3}(a) Enlargement of one adhesion site of the cell in the previous figures. Actin is shown in red, the cell edge in green, and the tops of the micropillars in blue. The lines represent the fitted circle (white) and ellipse (yellow). The arrows correspond to the measured forces (green) and the predicted directions (but not magnitudes) of the forces in the presence of isotropic ($\alpha=0$, white arrow) and anisotropic ($\alpha \ne 0$, yellow arrow) contractile stresses. Scalebar is 2 $\upmu$m. (b) Histogram (shown as a probability density) of $\theta_{\rm force}-\theta_{\rm shape}$ for isotropic (black) and anisotropic (orange) contractile stresses. Both the distributions are centered around $0^{\circ}$, the standard deviations are $60^{\circ}$ and $40^{\circ}$ for the isotropic and anisotropic models, respectively.}
\end{figure*}

In the following, we introduce a number of simplifications. As the orientation of the stress fibers varies only slightly along a single cellular arc [Fig. \ref{fig2}(a), and Figs. S2 and S3 in the Supplemental Material \cite{Sup}], we assume $\theta_{\rm SF}$ to be constant along each arc, but different, in general, from arc to arc. Furthermore, as all the arcs share the same bulk, we assume the bulk stresses $\sigma$ and $\alpha$ uniform throughout the cell. Under these assumptions a general solution of Eq. \eqref{shape} can be readily obtained. Taking $\bm{T}=(\cos\varphi,\sin\varphi)$, $\bm{N}=(-\sin\varphi,\cos\varphi)$, with $\varphi$ the orientation of the tangent vector $\bm{T}$ with respect to an axis perpendicular to the stress fibers [Fig.~\ref{fig2}(a)], and $\tan\varphi={\rm d}y/{\rm d}x$, with $(x,y)$ the position of the cell contour, yields:
\begin{multline}\label{eq_ellipse}
\frac{\sigma^{2}}{\gamma\lambda_{\min}^{2}}\left[(x-x_{c})\sin \theta_{\rm SF}-(y-y_{c})\cos\theta_{\rm SF}\right]^{2}\\
+\frac{\sigma^{2}}{\lambda_{\min}^{2}}\left[(x-x_{c})\cos \theta_{\rm SF}+(y-y_{c})\sin\theta_{\rm SF}\right]^{2}=1\;,
\end{multline}
where $\gamma=\sigma/(\sigma+\alpha)$ and $\lambda_{\min}$ is an integration constant related with cortical tension and whose physical interpretation will become clear later. Eq. \eqref{eq_ellipse} describes an ellipse of semiaxes $a=\sqrt{\gamma}\,\lambda_{\min}/\sigma$ and $b=\lambda_{\min}/\sigma$, centered at the point $(x_{c},y_{c})$ and whose major axis is parallel to the stress fibers, hence tilted by an angle $\theta_{\rm SF}$ with respect to the $x$ axis (Fig. \ref{fig2}). The dimensionless quantity $\gamma$ highlights the anisotropy of the forces acting on the cell contour. Thus, $\gamma=0$ corresponds to the case in which the directed forces outweigh the isotropic ones, whereas $\gamma=1$ reflects the purely isotropic case. Further details can be found in \cite{Sup}.

The key prediction of our model is illustrated in Fig.~\ref{fig2}(b), where we have fitted the contour of the same cell shown in Fig.~\ref{fig1}(a) with ellipses. More examples are shown in Fig.~S2 and S3 in the Supplemental Material \cite{Sup}. Whereas large variations in the circles' radii were required in Fig.~\ref{fig1}(a), a {\em unique} ellipse ($\gamma=0.52$, $\lambda_{\min}/\sigma=13.4\,\upmu{\rm m}$) faithfully describes all the arcs in the cell. The directions of the major axes were fixed to be parallel to the local orientations of the stress fibers in the fit. To test the accuracy of this latter choice, we fitted unconstrained and independent ellipses to all cellular arcs in our database. The distribution of the difference between the orientation $\theta_{\rm ellipse}$ of the major axis of the fitted ellipse and the measured orientation $\theta_{\rm SF}$ of the stress fibers is shown in Fig.~\ref{fig2}(c). The distribution peaks at $0^{\circ}$ and has a width of $36^{\circ}$, demonstrating that the orientation of the ellipses is parallel, on average, to the local orientation of the stress fibers as predicted by our model.

Eq.~\eqref{shape} further allows to analytically calculate the cortical tension $\lambda$. Namely,
\begin{equation}\label{eq_tension}
\lambda(\varphi) = \lambda_{\rm min} \sqrt{\frac{1 + \tan^{2}(\varphi)}{1 + \gamma \tan^{2}(\varphi)}}\;.
\end{equation}
The function $\lambda$ attains its minimum value at the point along the cellular arc where $\varphi=0$ and $\lambda(0)=\lambda_{\min}$. Here, the cortical tension has no contribution from the directed stress (i.e., $\bm{n}\cdot\bm{T}=0$), thus $\lambda_{\min}$ represents the minimal tension withstood by the cortical actin. Although the latter could, in principle, be arc-dependent, for instance in the presence of substantial variations in the actin densities \cite{Bischofs2008}, here we approximate $\lambda_{\rm min}$ as a constant. Thus  $\sigma$, $\alpha$ and $\lambda_{\min}$ represent the material parameters of our model.

Eqs.~\eqref{eq_ellipse} and \eqref{eq_tension} are combined to predict the traction force exerted by the cell at a specific adhesion site by adding the cortical tension $\lambda \bm{T}$ along the two cellular arcs joining at the adhesion site. We emphasize that this analysis yields information on cellular forces solely based on the analysis of cell shape. For example, the direction of the traction forces is calculated without additional fitting parameters. We compare the result with the direction of the traction force measured with a micropillar array technology \cite{Tan2003,Trichet2012a,VanHoorn2014}. An example is shown in Fig.~\ref{fig3}(a) for one of the adhesion points of the cell in Fig.~\ref{fig2}(b); more examples are shown in Figs. S2 and S3 in the Supplemental Material \cite{Sup}. The arrows mark the direction of the measured traction force (green) and that calculated by approximating the cell shape with ellipses (yellow). As a comparison, Fig.~\ref{fig3}(a) also shows a prediction based on circles from the isotropic model (white) \cite{Bischofs2008,Bischofs2009}.

In Fig.~\ref{fig3}(b), we show the distribution of the difference $\theta_{\rm{force}}-\theta_{\rm{shape}}$ between the measured orientation of the traction forces and that calculated from our model, across the entire cell population. The predicted distribution is centered at $0^{\circ}$ and has a width of $40^{\circ}$. As a comparison, we also plot the result for the isotropic model, which displays a larger standard deviation of about $60^{\circ}$. This shows that not only cell shape, but also adhesion forces are profoundly affected by the anisotropy of the cytoskeleton. 

\begin{table}[t]
\begin{ruledtabular}
\begin{tabular}{cccc}
$\gamma$ & $\lambda_{\min}\,({\rm nN})$ & $\sigma\,({\rm nN}/\upmu{\rm m})$ & $\alpha\,({\rm nN}/\upmu{\rm m})$ \\
\hline
$0.33 \pm 0.20$ & $7.6 \pm 5.6$ & $0.87 \pm 0.70$ & $1.7 \pm 1.7$
\end{tabular}	
\end{ruledtabular}
\caption{\label{tab_parameters}Survey of the average material parameters in a sample of 285 fibroblastoid and epithelioid cells.}
\end{table}

Finally, our model allows us to obtain quantitative information on the relative magnitude of isotropic and anisotropic stresses. In Tables~\ref{tab_parameters} and S1 \cite{Sup} we report a survey of the material parameters over a sample of 285 cells. Despite the large variability among the cell population, the directed stress $\alpha$ is consistently larger than the isotropic stress $\sigma$, reflecting the high anisotropy of the adherent cell types used here.

In conclusion, we have investigated the geometrical and mechanical properties of adherent cells characterized by an anisotropic actin cytoskeleton, by combining experiments on micropillar arrays with simple mechanical modeling. We have predicted and tested that the shape of the cell edge consists of arcs that are described by a unique ellipse, whose major axis is parallel to the orientation of the stress fibers. The model allowed us to obtain quantitative information on the values of the isotropic and anisotropic contractility of cells. In the future, we plan to use our model in combination with experiments on micropatterns (see, e.g., Refs. \cite{Thery:2006,Thery:2010}), where cellular shape can be controlled, thus allowing higher reproducibility of the results and more systematic statistical analysis of the data.

\begin{acknowledgments}
This work was supported by funds from the Netherlands Organisation for Scientific Research (NWO/OCW), as part of the Frontiers of Nanoscience program (L.G.), the Netherlands Organization for Scientific Research (NWO-FOM) within the program on Barriers in the Brain (W.P. and T.S.; No. FOM L1714M), the Netherlands Organization for Scientific Research (NWO-ALW) within the Innovational Research Incentives Scheme Vidi Cross-divisional 2010 ALW (R.M.H.M.; No. 864.10.009), and the Leiden/Huygens fellowship (K.S.).
\end{acknowledgments}

%

\end{document}